\documentclass[conference,10pt]{IEEEtran}
\IEEEoverridecommandlockouts
\usepackage{cite}
\usepackage{amsmath,amssymb,amsfonts}
\usepackage{algorithmic}
\usepackage{graphicx}
\usepackage{textcomp}
\usepackage{xcolor}
\usepackage{multirow}
\def\BibTeX{{\rm B\kern-.05em{\sc i\kern-.025em b}\kern-.08em
		T\kern-.1667em\lower.7ex\hbox{E}\kern-.125emX}}
	
\usepackage[letterpaper, portrait, margin=0.7in, bmargin=0.95in]{geometry}

\usepackage{fancyhdr}
\fancypagestyle{firststyle}{\fancyhf{}
	\fancyhead[L]{D. Shakya, D. Chizhik, J. Du, R. A. Valenzuela, and T. S. Rappaport ``Dense Urban Outdoor-Indoor Coverage from 3.5 to 28 GHz",\textit{ in ICC 2022 - 2022 IEEE International Conference on Communications, }Seoul, South Korea, May 2022, pp. 1--6}
}

\begin{document}
	\bstctlcite{IEEEtran:BSTcontrol}
	\title{Dense Urban Outdoor-Indoor Coverage from 3.5 to 28 GHz
		\thanks{This work is supported by NOKIA Bell Labs, the NYU WIRELESS Industrial Affiliates program, and National Science Foundation (NSF) Research Grants: 1909206 and 2037845}
	}
	
	\author{\IEEEauthorblockN{Dipankar Shakya$^{\dagger*}$, Dmitry Chizhik$^{\dagger}$, Jinfeng Du$^{\dagger}$, Reinaldo A. Valenzuela$^{\dagger}$, and Theodore S. Rappaport$^{*}$}
		\IEEEauthorblockA{$^{\dagger}$NOKIA Bell Labs, New Providence, NJ-07974,\{dmitry.chizhik, jinfeng.du, reinaldo.valenzuela\}@nokia-bell-labs.com\\
			$^{*}$NYU WIRELESS, NYU Tandon School of Engineering, Brooklyn, NY-11201,\{dshakya, tsr\}@nyu.edu}
	}

	\maketitle
	
	\linespread{1.05}
	
	\thispagestyle{firststyle}
	
	\begin{abstract}
		In the US, people spend 87\% of their time indoors and have an average of four connected devices per person (in 2020). As such, providing indoor coverage has always been a challenge but becomes even more difficult as carrier frequencies increase to mmWave and beyond. 
		This paper investigates the outdoor and outdoor-indoor coverage of an urban network comparing globally standardized building penetration models and implementing models to corresponding scenarios. The glass used in windows of buildings in the grid  plays a pivotal role in determining the outdoor-to-indoor propagation loss. For 28 GHz with 1 W/polarization transmit power in the urban street grid, the downlink data rates for 90\% of outdoor users are estimated at over 250 Mbps. In contrast, 15\% of indoor users are estimated to be in outage, with SNR $<-3$ dB when base stations are 400 m apart with one-fifth of the buildings imposing high penetration loss ($\sim$ 35 dB). At  3.5 GHz, base stations may 
		achieve over 250 Mbps for 90\% indoor users  if 400 MHz bandwidth  with 100 W/polarization transmit power is available. The methods and models presented can be used to facilitate decisions regarding the density and transmit power required to provide high data rates to majority users in urban centers.  
	\end{abstract}
	
	
	\section{Background}
	The US Environmental Protection Agency, through the National Human Activity Pattern Survey in 2001, determined that in the US people spend 87\% of their time indoors with an additional 6\% inside vehicles \cite{klepeis2001nature}. Additionally, the Cisco 2018-2023 annual internet report for the US indicates an average of four connected devices per person in 2020 \cite{cisco2020ir}. Thus, it becomes imperative to know how much signal coverage and data rate can be provided to indoor users and devices, especially for urban networks with building structures that may not favor penetration of cellular signals from outdoors.
	
	The outdoor-to-indoor (O2I) or building penetration loss (BPL) has been a long standing problem for wireless communications, and significant channel measurement and modeling work was conducted when the cellular radio industry was in its infancy \cite{Rappaport1994apm}. Measurement campaigns have been conducted to characterize wireless propagation\cite{Larsson2014eucap, Bas2019twc, Du2018wcl, ITUR2346} and behavior of materials\cite{Zhao2013icc,Ryan2017iccw} for O2I coverage. The O2I measurement campaign in \cite{Larsson2014eucap} observes that O2I coverage is limited by attenuation/excess loss imposed along the signal path. Further, the location and type of windows are observed as important factors that might affect cellular service deployment. Measurements performed in \cite{Bas2019twc} note that the strongest paths received indoors from an outdoor transmitter have normal incidence into the building or were reflections from window frames. \cite{Zhao2013icc} records reflection and penetration measurements for common building materials and windows and suggests O2I communications is difficult at 28 GHz. Further, \cite{Ryan2017iccw} shows influence of antenna polarization on penetration loss at 73 GHz as some indoor materials, such as closet doors, exhibited lower loss with cross-polarized antennas. \cite{Du2018wcl} measures penetration loss in sub-urban houses emulating a customer premises equipment (CPE) placed outside or 1.5 m inside a street facing window to relay signals indoors; An additional loss of 9 dB for house with low-loss wooden exterior and 17 dB for house with foil back insulation and low-emissivity (low-e) windows is observed when moving the CPE indoors. 
	
	In the literature, buildings are essentially categorized into two groups: low-loss and high-loss, based on metrics relating to the building thermal efficiency. The 3GPP recommendation for BPL follows from the proportion of two materials, primarily concrete and glass, found on the building surface area and groups buildings as high-loss or low-loss \cite{3GPPTR38901}. Penetration loss of each material is modeled as a linear function of frequency, based on formulations in \cite{Semaan2014gc}. Similarly, the ITU-R recommendation classifies buildings as traditional (low penetration loss) or thermally efficient (high penetration loss), as per the thermal transmittance (U-value) to determine building entry loss (BEL) \cite{ITUR2109}. The BEL model is used in addition to path loss models provided in \cite{ITUR1411}.
	
	For both models, the type of glass windows used in the building plays a defining role in determining the extent of loss experienced by the signal. The Commercial Buildings Energy Consumption Survey (CBECS) \cite{CBECS2018} and Residential Energy Consumption Survey (RECS) \cite{RECS2015} surveys list the window types in common use as single-pane, double-pane, and triple-pane. The glass used for the windows may be clear, tinted, or coated for low-e. Single-pane windows are found to be clear glass while double-pane windows may have a thin metallic coating (typically silver) for low-e, alloys added for tinting and infrared (IR) reflection, or use clear glass. Triple paned windows typically have two layers of low-e coating. In \cite{Zhao2013icc}, authors report a penetration loss of 3.9 dB through clear glass at 28 GHz, while tinted glass attenuated the signal by 40 dB agreeing with observations in \cite{Larsson2014eucap}. Likewise, in \cite{TIP2019} 27.4 dB loss was measured for penetration through double-pane low-e glass at 28 GHz. 
	To allow wireless signals to pass through the low-e metallic coating in windows, \cite{Widenberg2002lu} simulates frequency selective structures implemented as periodically perforated apertures on the metal coating.
	
	Based on CBECS 2012 and RECS 2005 reports, \cite{3GPP163408} defines single and double-pane clear glass as low-loss, double-pane low-e glass as medium-loss, and tinted, IR reflective (IRR), and triple-pane glass as high-loss. The market share distribution for windows is found at 85\% low-loss, 5\% medium-loss, and 10\% high-loss in 2010; Thereafter, \cite{3GPP163408} forecasts minor increase in percentages of medium and high loss windows until 2025 using new window sales data (Table 5 and Figure 6 in \cite{3GPP163408}). Thus, \cite{3GPP163408} recommends using the 3GPP low BPL model with 80-90\% probability in urban scenarios; It justifies the low-loss BPL allocation in proportion to the distribution of low-loss windows (i.e. $\geq$80\%) stating that street clutter is avoided on upper buildings floors and users are more likely to find LOS paths. 
	
	The indoor coverage is described in the general building penetration model \cite{Semaan2014gc} showing signals reaching the indoor user equipment (UE) from four surrounding walls along the shortest path and then selecting the lowest path loss signal for received power. \cite{Sheikh2017iwcmc} extends the model by adding the received powers from the four directions and additionally considers a direct path from the BS to the UE for a better description of the indoor coverage. The signal paths may reach the exterior of the building from all four directions due to around the corner propagation owing to diffraction, scattering and reflection off objects at street corners such as lampposts\cite{Lu2014tap, Rappaport2017tap,DuD2020tap}.Key findings include:
	\begin{itemize}
		\item Distribution of BPL defines the outage ratio of indoor users: 15\% outage if only 20\% buildings has high loss, and 60\% outage if all buildings were high loss.
		\item With 400 m inter-site distance (ISD) in dense urban street grid, 100 W transmit power per polarization  results in interference limited system, even at 28 GHz.  
	\end{itemize}
	
	
	\section{The Urban Environment Setup}
\begin{figure}[t]
		\centering
		\includegraphics[width=0.34\textwidth]{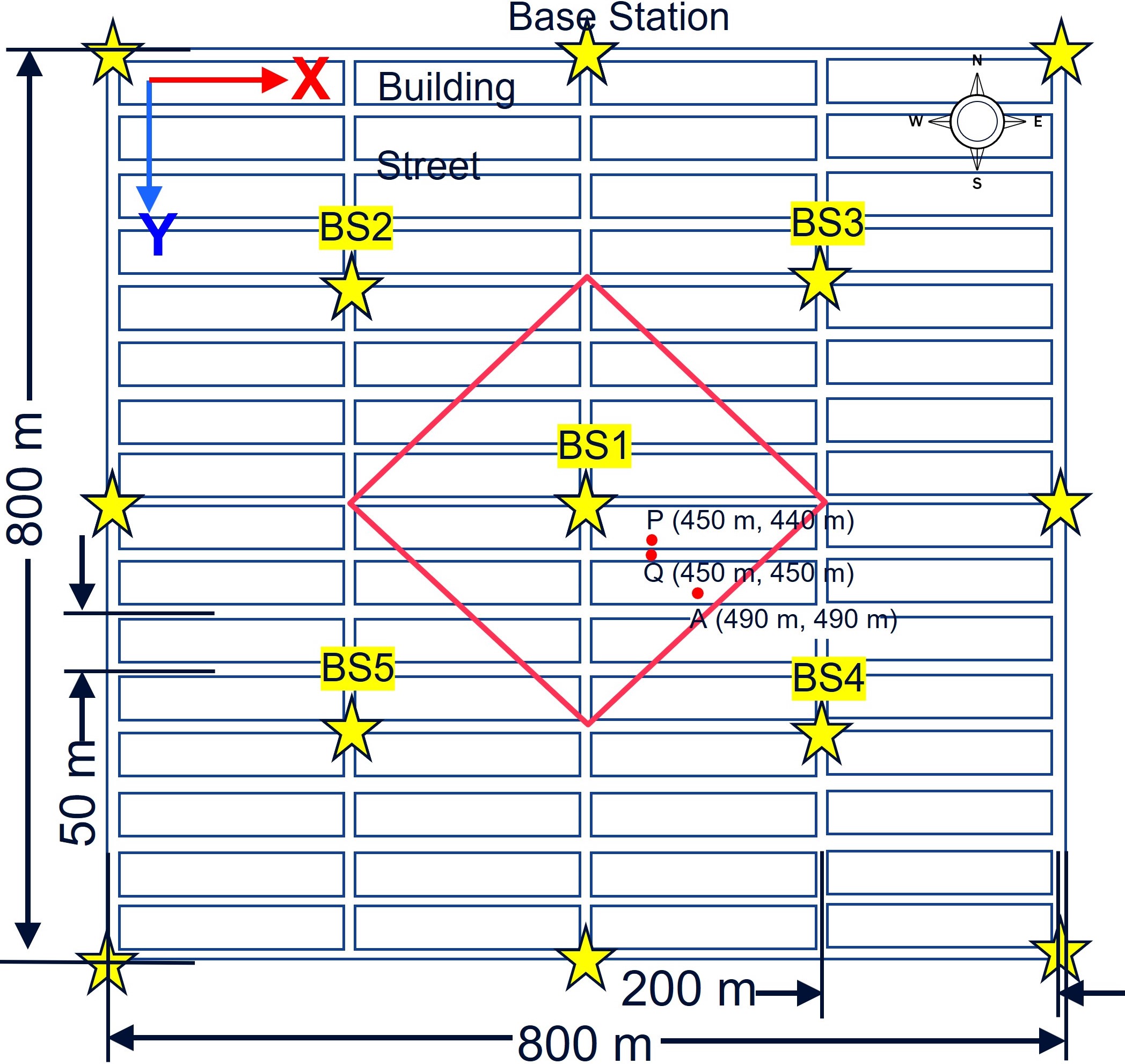}
		\caption{Urban grid layout for analysis. The red diamond is a boundary demarcation for locations that are statistically representative of any BS on the grid. Starts represent the BSs. The red dots A, P, and Q represent arbitrary coordinates within the boundary. Coordinate (0,0) is at the top-left corner.}
		\label{fig:Layout}
		\vspace{-0.5 cm}
	\end{figure}

	The urban environment is setup as a street-grid with blocks of buildings and streets between them. A rectangular grid layout of 800 m × 800 m with uniform blocks of 200 m × 50 m including buildings and streets, as shown in Fig.\ref{fig:Layout}, is considered for the simulations.
	
	Within the layout, base stations are initially placed at street intersections with an ISD of 400 m, resulting in a BS density of 12 BSs per sq. km. Each base station is assumed mounted on the rooftop edge of a corner building at a street intersection resulting a BS antenna height of 25 m. For the urban scenario, the corner buildings with base-stations are not assumed to be the tallest buildings on the map, but of intermediate height. 
	
\begin{figure}[t]
		\centering
		\includegraphics[width=0.44\textwidth]{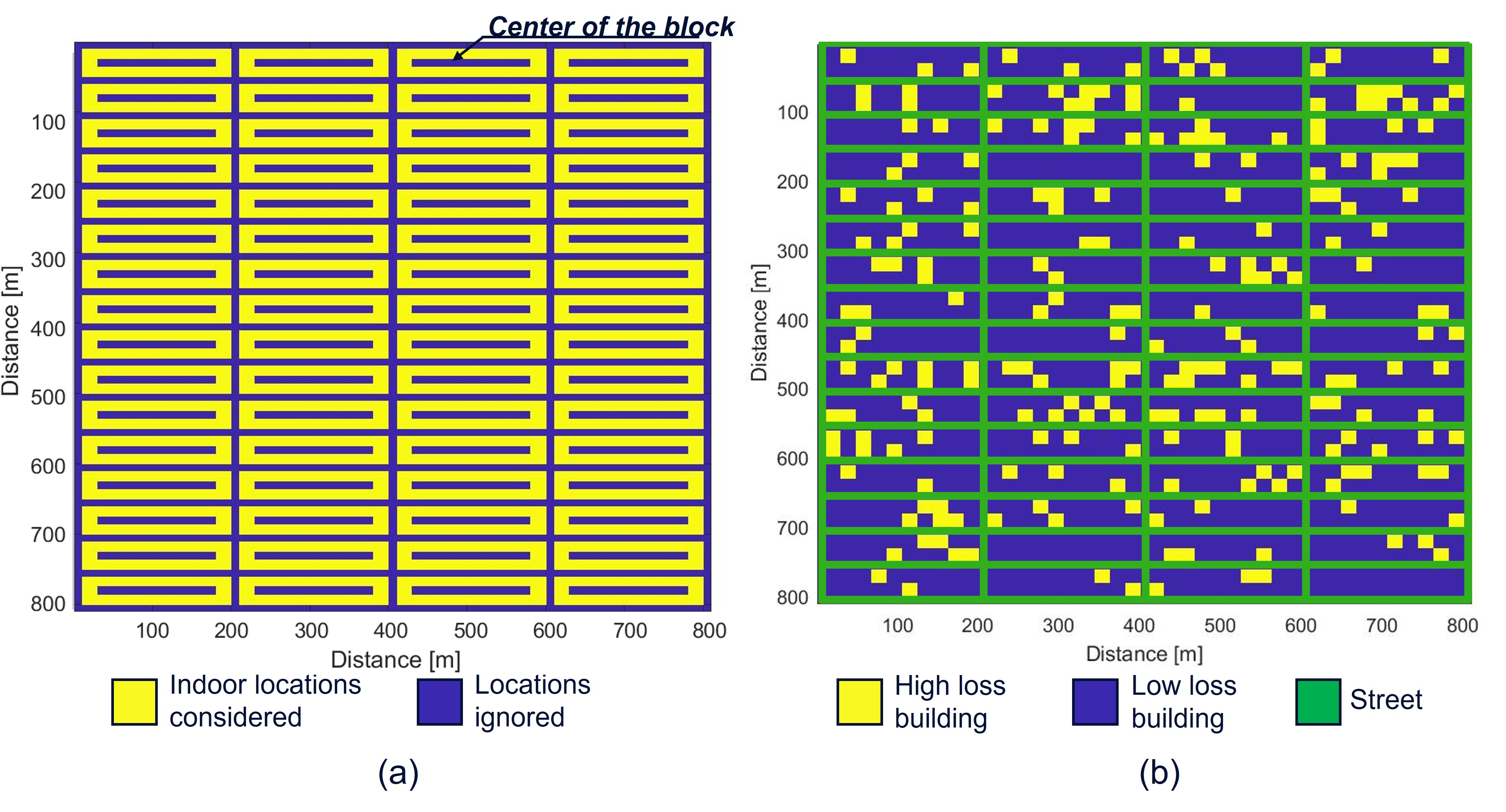}
		\caption{(a) Building blocks on the street grid highlighted in yellow. For UE locations indoors, the isolated blue strips at the center of building blocks are ignored.(b) Randomly generated distribution of buildings that present a high penetration loss (yellow) on the street grid.}
		\label{fig:gridLogic}
		\vspace{-0.9 cm}
	\end{figure}

	The UEs located on the same street as a BS are considered LOS, while those on streets without a BS are considered NLOS. The signal is assumed to reach NLOS UEs via diffraction and scattering at corners of streets having a BS and over the rooftops.  
	The buildings take up 190 m $\times$ 40 m of the 200 m $\times$ 50 m block and may exhibit either low or high penetration loss. The final simulations are conducted with 80\% low BPL and 20\% high BPL considering typical concrete multistory buildings with good insulation\cite{3GPP163408}. A building, in this simulation, is defined as 19 m × 20 m rectangle within the 190 m $\times$ 40 m building-block in the grid. Additionally, the center of city blocks are observed to be generally unoccupied and filled with vegetation, or have alleyways and gaps left for windows. When considering indoor UE locations for statistical analysis, the locations within a 150 m $\times$ 10 m area at the center of the block of buildings are ignored, as illustrated in Fig.~\ref{fig:gridLogic}(a). The low-loss and high-loss buildings are randomly distributed throughout the grid for simulation. A resulting distribution may appear as shown in Fig.~\ref{fig:gridLogic}(b).
	
	The locations within the bounded region shown in Fig.~\ref{fig:Layout} are considered for statistical analysis as the bound includes interference effects uniformly from all directions. As a contrary example, if the BS at the top-right corner is considered, three-fourths of the influence from neighboring BSs would be overlooked due to the limited grid size and statistical results could vary. The diamond shape assures replicability of results across a grid with any number of BSs and same BS density as the bound would not overlap with bounds for other BSs.
	
	\section{Propagation Loss Models}
	
	\subsection{Outdoor Propagation Models}
	\subsubsection{Same Street path gain}
	The propagation along streets having a BS is defined using models in \cite{DuD2020tap} for same-street coverage with the antenna mounted on the rooftop edge of a corner building at a street intersection. The models are based on measurements conducted from buildings to streets in Manhattan, New York, USA and Valparaíso, Chile. The path gain (PG) model based on distance (d [m]) is formulated with a 1-m intercept (A [dB]), distance exponent (n), and RMS error ($\sigma$ [dB]) as follows:
	\begin{equation}
	\label{eq:ss_pg}
	PG_{same-street}\ \ [dB]=A - 10n\log_{10}(d)+\mathcal{N}(0,\sigma),
	\end{equation}
	\begin{equation*}
	\begin{split}
	A=-35, n=3.56,\sigma =7.1\ dB
	\end{split}	
	\end{equation*}
	
	\subsubsection{Around the corner}
	The signal from a BS can provide coverage into perpendicular streets through diffraction around corner buildings, scattering, and/or reflection at street intersections. The model for around-the-corner propagation, formulated in \cite{DuD2020tap}, employs an empirical corner loss obtained from real world measurements.
	\begin{equation}
	\label{eq:crnrPG}
	\begin{split}
	\hspace{-3mm} PG_{d}(x)\text{ [dB] }{=}
	\begin{cases}
	P_{1} {-} 10n\log_{10}(x),\indent \indent 1<x<d_{c}\\
	P_{1} {-} \Delta {-}5n\log_{10}(d_{c}(x{-}d_{c})x),\text{ }x{>}d_{c} 
	\end{cases} \hspace{-3mm}
	\end{split}
	\end{equation}
	
	where $d_c$ is the distance from the BS to the street corner, $P_1$ is the 1-m distance intercept, n is the distance exponent and $\Delta>0$ is the empirical “corner loss”.
	
	\subsection{O2I penetration loss}
	3GPP TR 38.901 specifies the O2I penetration loss as the sum of four constituents as presented in \eqref{eq:BPL}.
	\begin{equation}
	\label{eq:BPL}
	\begin{split}
	PL\text{ [dB] }= PL_{b} + PL_{tw} + PL_{in} + \mathcal{N}(0,\sigma_{p})
	\end{split}
	\end{equation}
	
	Here, $PL_{b}$ [dB] is the basic path loss from BS to UE governed by scenario of implementation and obtained from the models provided in \cite{3GPPTR38901}. For analysis, the Urban macrocell (UMa) scenario is chosen, owing to the 25 m BS height and 400 m ISD. $PL_{tw}$ [dB] in \eqref{eq:BPL} represents BPL and is obtained as the weighted average of the loss from different materials on the building exterior, as in \eqref{eq:PLtw}. The weights, $p_{i}$, are obtained as the proportion of the surface area occupied by the material on the building wall and $PL_{in}$ [dB] is the depth-dependent indoor path loss ($PL_{in}= 0.5d_{2D-in}$). $d_{2D-in}$ [m] is the shortest distance from the surrounding wall to the indoor location and $\sigma_{p}$ is the RMS error in the BPL.
	\begin{equation}
	\label{eq:PLtw}
	\begin{split}
	PL_{tw} [dB]= PL_{npi}-10\log_{10}\sum_{i=1}^{N}p_{i}\times 10^{\frac{L_{material\_i}}{-10}}
	\end{split}
	\end{equation}
	
	Here, $PL_{npi}=5$ dB is the loss from non-perpendicular incidence~\cite{3GPPTR38901}. For general use across different propagation scenarios, 3GPP gives two variants of the model in \eqref{eq:PLtw} as:\\
	\textbf{Low loss model}:- material\_1: glass with $p_{1}=30 \%$ \& material\_2: concrete with $p_{2}=70 \%$ \\
	\textbf{High loss model}:- material\_1: IRR glass with $p_{1}=70 \%$ \& material\_2: concrete with $p_{2}=30 \%$ 
	
	The materials showcase a frequency dependent loss as described below (Here $f$ is in GHz):
	\begin{itemize}
		\item Standard multi-pane glass: $L_{\mathrm{glass}}$ [dB] $=2+0.2f$
		\item IRR glass: $L_{\mathrm{IRRglass}}$ [dB] $=23+0.3f$
		\item Concrete: $L_{\mathrm{concrete}}$ [dB]$= 5+4f$
	\end{itemize}
	
\begin{figure}[htbp]
		\centering
		\includegraphics[width=0.42\textwidth]{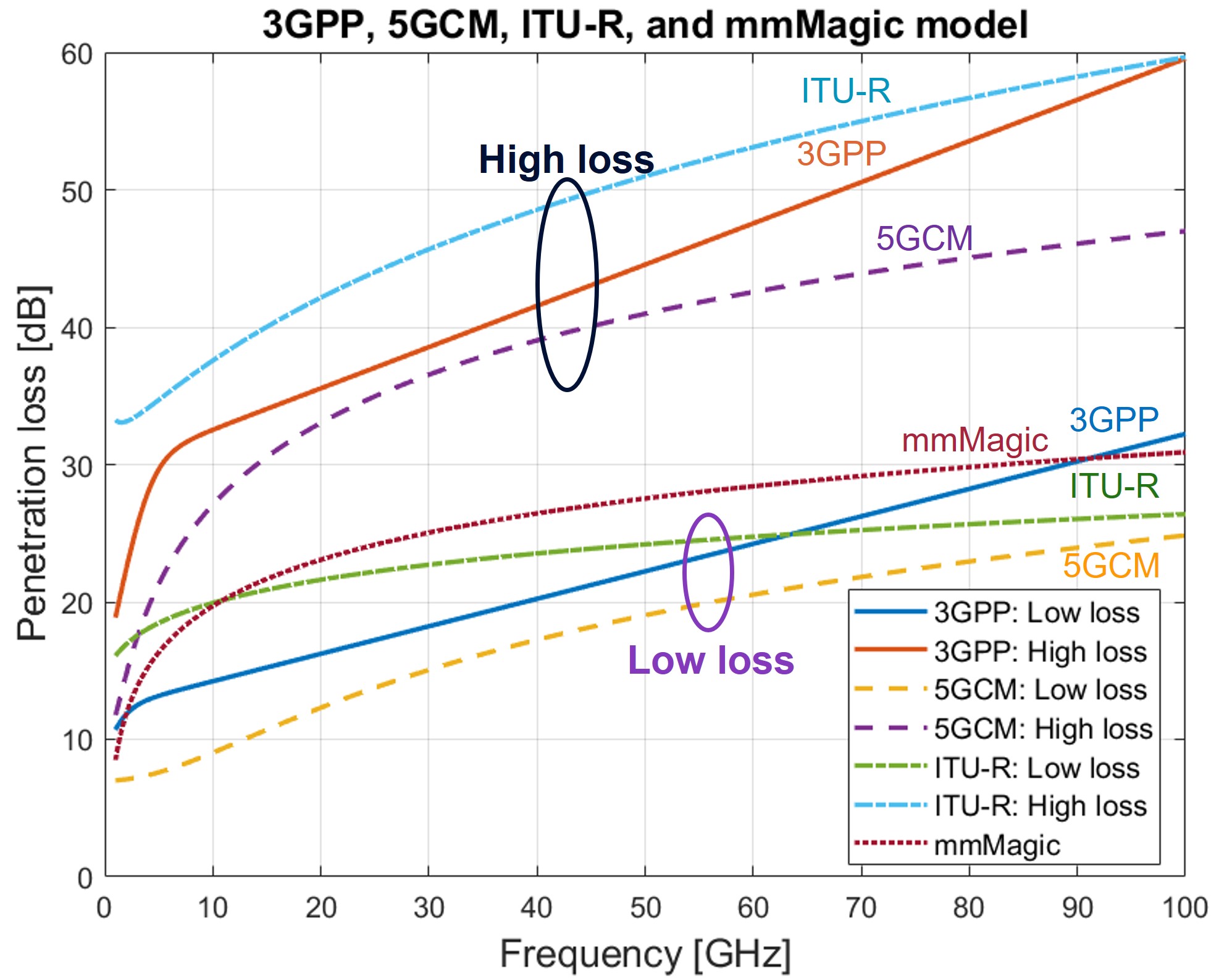}
		\caption{Building penetration loss models up to 100 GHz. Each recommendation effectively has a high and low loss classification based on materials used in the building construction.}
		\label{fig:BPL3}
		\vspace{-0.5 em}
	\end{figure}

	\textbf{Comparison of 3GPP, 5GCM, ITU-R, and mmMagic models:}
	The 5GCM, ITU-R, and mmMagic also provide recommendations for BPL up to 100 GHz \cite{5GCM2016wp,Rappaport2017tap2,ITUR2109}. The results of the penetration loss models are plotted in Fig.\ref{fig:BPL3} and essentially classify BPL as high-loss or low-loss. \textcolor{black}{The 5GCM model provides an alternative two-parameter model with simpler parameter variations that closely fits the 3GPP model results for perpendicular incidence~\cite{5GCM2016wp,Rappaport2017tap2}, 
	}
	\textcolor{black}{\[BPL[dB]=10\log_{10}(A+Bf^{2}),\]}
	where $f$ is carrier frequency in GHz, ($A{=}5,B{=}0.03$) for low-loss building types and ($A{=}10,B{=}5$) for high-loss. 
	The ITU-R models, represented by the dash-dot lines, are generated from a model with nine parameters based on building type and have specific parameter values for “thermally-efficient” (high-loss) and “traditional” (low-loss) building types, derived from measurements collated in \cite{ITUR2346}. Monte-Carlo simulations may then be performed taking the mean of the results as the BEL at each frequency\cite{ITUR2109}. \textcolor{black}{The mmMagic provides a single model for O2I penetration without classification, resulting in values typically higher than the other low-loss models\cite{Jaeckel2016ewc,Rappaport2017tap2}. Among the models, The 5GCM model may be easily tuned to measured data, and more data will be needed for the industry to use the most accurate model for future deployments and analyses.}
	
	\textcolor{black}{The 3GPP O2I penetration loss agrees with the trend of the 5GCM and ITU-R models. It considers the non-perpendicular incidence and losses from individual materials on the building exterior, while fitting closely with measured data. Hence, considering the selectivity in window types for this work, the 3GPP O2I penetration loss model is employed.}

	\subsection{Spatial Correlation for Shadow Fading}
	The drop-based modeling implemented in this project requires independent shadow fading for each UE point simulated. However, when generating the coverage map of signal-to-noise ratio (SNR) in the entire urban street grid at one time-instant, the shadow fading between nearby points in space would be correlated by virtue of existing in the same environment. 3GPP \cite{3GPPTR38901} and ITU-R \cite{ITUR1816} recommend using an exponential function to describe the autocorrelation of shadow fading. 
	The spatial filter is applied over iid random values of shadow fading for propagation in LOS, NLOS, or indoors and is implemented as a 2D spatial filter described in \cite{Ju2019gc, Ju2018vtc}.
	
	\subsection{Sector Interference}
	The interference between sectors is simulated using a key-hole radiation pattern, as described in \cite{Wang2017sen}, for the BS antenna. The main lobe is assigned a gain equal to transmit antenna gain, $G_{tx}$ [dB], and half-power beamwidth (HPBW) of 10 degrees \cite{DuD2020tap}. Using the HPBW and $G_{tx}$, the side-lobe gain ($G_S$) can be calculated using Equation (13) in \cite{Wang2017sen}. Assuming a frequency reuse of 1, the interference from the other sectors is calculated as the sum of powers received at the UE location with a $G_{tx}$ equal to $G_S$. Thus, for four sectors implemented at the BS, three other sectors would interfere in equal amounts with an antenna gain of $G_S$.
	
	\section{Simulation Parameters}
	The path gain for signal propagation on the streets of the grid is calculated for each BS in multiple steps. Considering the center BS in the 800 m $\times$ 800 m grid, first, the path gains along the same street as the BS in East-West (E-W) and North-South (N-S) directions are calculated using \eqref{eq:ss_pg}. Thereafter, from the street with the BS, path gains along perpendicular streets \textcolor{black}{are determined considering propagation around-the corner and along a direct path reaching the NLOS UE over rooftops and through buildings, using \eqref{eq:crnrPG} and NLOS UMa models in \cite{3GPPTR38901}, respectively.} The same process is repeated for every BS with coverage at each street provided by a one-turn propagation into the perpendicular street from the street with the BS. Finally, considering all the BSs, the maximum path gain at each location is selected for the signal reaching the user. The signals reaching the location from all other BSs are considered as interference to the strongest signal.
	
\begin{figure}[htbp]
		\centering
		\includegraphics[width=0.46\textwidth]{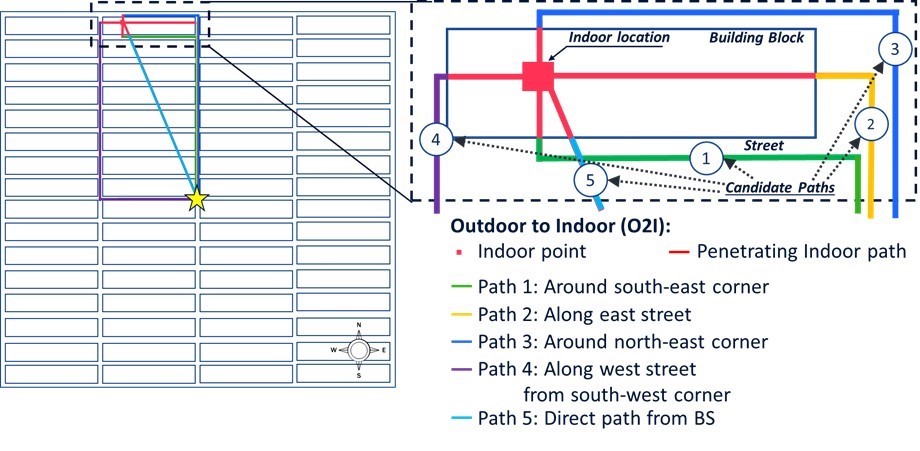}
		\caption{Candidate paths reaching the indoor UE from the surrounding four walls (paths 1-4) and over-clutter-tops (path 5). Paths with two or more turns around corners are neglected.}
		\label{fig:Paths}
		\vspace{-0.5 em}
	\end{figure}

	With the path-gain at each street location obtained, the indoor propagation is modeled considering paths penetrating indoors from the streets through all four surrounding walls, as illustrated in Fig.\ref{fig:Paths}. In addition to paths 1-4, a  NLOS over-clutter-tops path from the rooftop mounted BS to the indoor user is indicated as path 5. The power from path 5, though weak, is added as part of the total received power. A threshold is set for the maximum distance (10 m) the direct path travels indoor, beyond which it is set to the threshold value. This value is set assuming that the indoor user is always within a threshold distance from a wall with a window. The power from the five candidate paths are added to obtain the total received power.
	
	Each location on the 800 m $\times$ 800 m grid is spaced 1 m apart. For the evaluated PGs at each location, shadow fading corresponding to LOS, NLOS, or indoors is independently. Only for generating the SNR coverage map, the spatial correlation 2D filter in \cite{Ju2018vtc} is convolved with the shadow fading values at each location. The correlation distances for LOS, NLOS, or indoor scenarios are specified in \cite{3GPPTR38901} and the RMS error values are obtained from \cite{DuD2020tap}. Table \ref{tab:Sim_param} summarizes the parameters used in the simulations. 
	
	\begin{table}[ht]
		\caption{Parameter values used for the simulations}
		\label{tab:Sim_param}
		\begin{center}
			\begin{tabular}{|l|l|}
				\hline
				\textbf{Parameter} &      \textbf{Value} \\
				\hline
				Carrier frequency ($f_c$) &     3.5/7/14/28 GHz \\
				\hline
				\multicolumn{ 1}{|l|}{\multirow{2}{*}{BS transmit power ($P_{tx}$)}} & 50 dBm/pol., \\
				\multicolumn{ 1}{|l|}{} & 30 dBm/pol. at 28 GHz \\
				\hline
				Transmission bandwidth (BW) &    400 MHz \\
				\hline
				Number of polarizations (Pol) &          2 \\
				\hline
				BS antenna gain ($G_{tx}$) &     26 dBi \\
				\hline
				Height of rooftop BS antenna ($h_t$) &       22 m \\
				\hline
				\multicolumn{ 1}{|l|}{\multirow{2}{*}{UE antenna gain ($G_{ue}$)}} & 12 dBi (indoor CPE) \\
				\cline{2-2}
				\multicolumn{ 1}{|l|}{} & 6 dBi (outdoor streets) \\
				\hline
				Height of UE antenna ($h_{ue}$) &      1.5 m \\
				\hline
				\multicolumn{ 1}{|l|}{\multirow{2}{*}{Antenna gain degradation (M)}} & 2 dB (LOS) \\
				\cline{2-2}
				\multicolumn{ 1}{|l|}{} & 5 dB (NLOS) \\
				\hline
				Noise figure (NF) &       9 dB \\
				\hline
				Minimum detectable SNR/SINR &      -6 dB \\
				\hline
				Implementation penalty &       3 dB \\
				\hline
				Maximum depth for path 5 (Fig.\ref{fig:Paths}) &       10 m \\
				\hline
				BS antenna gain in adjacent sectors ($G_S$) &		4 dBi \\
				\hline
				
			\end{tabular}  
		\end{center}
		\vspace{-2.0 em}
	\end{table}
	
	\section{Results and discussion}
	
\begin{figure}[htbp]
		\centering
		\includegraphics[width=0.33\textwidth]{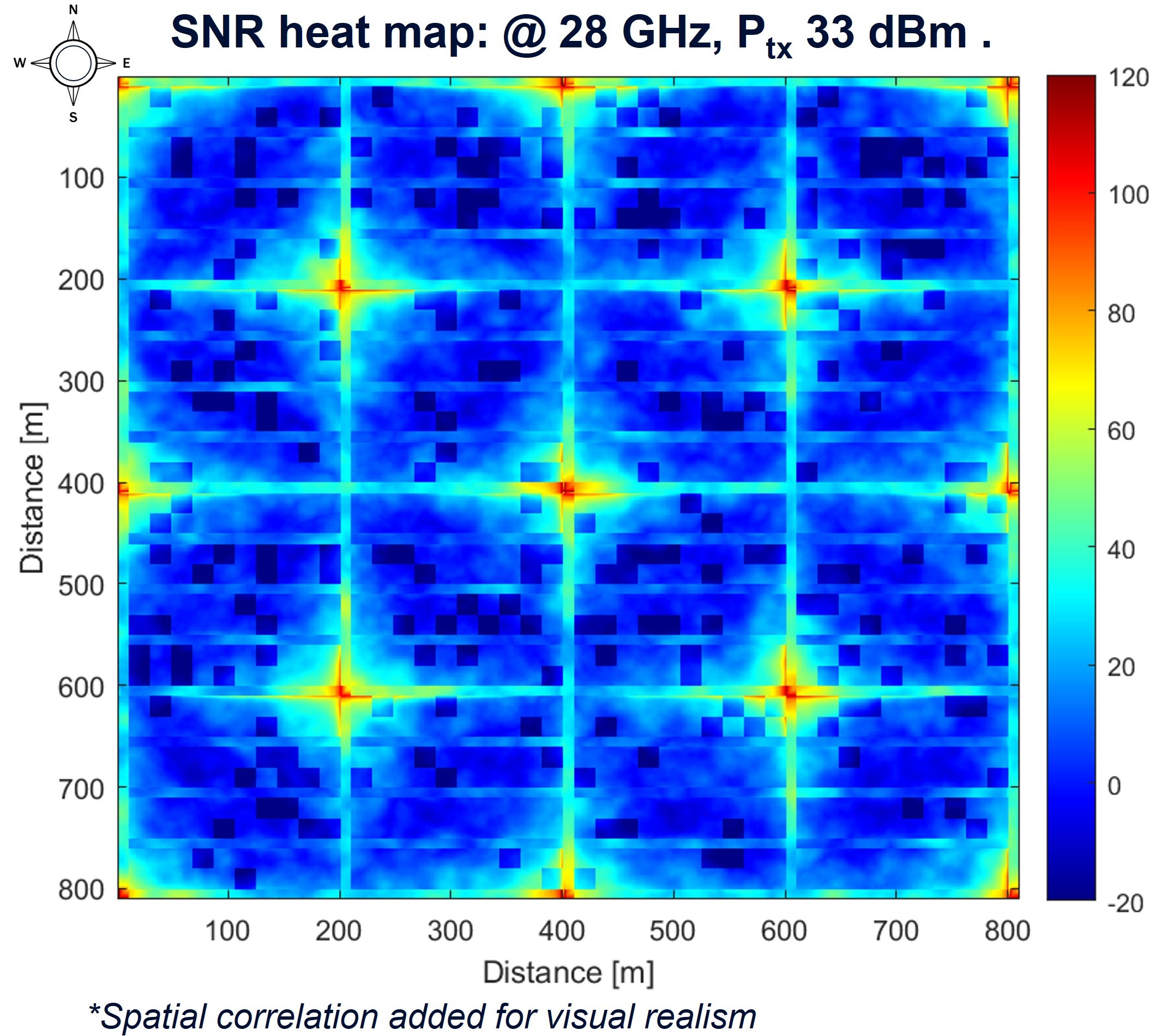}
		\caption{Downlink SNR coverage heat map at 28 GHz, 12 BS/ sq.km, 400 m ISD, and 30 dBm/pol. transmit power (Table \ref{tab:Sim_param}). The darker patches represent the buildings in the grid with high-loss exteriors.}
		\label{fig:heat}
		\vspace{-0.5 em}
	\end{figure}

\begin{figure}[ht]
		\centering
		\includegraphics[width=0.33\textwidth]{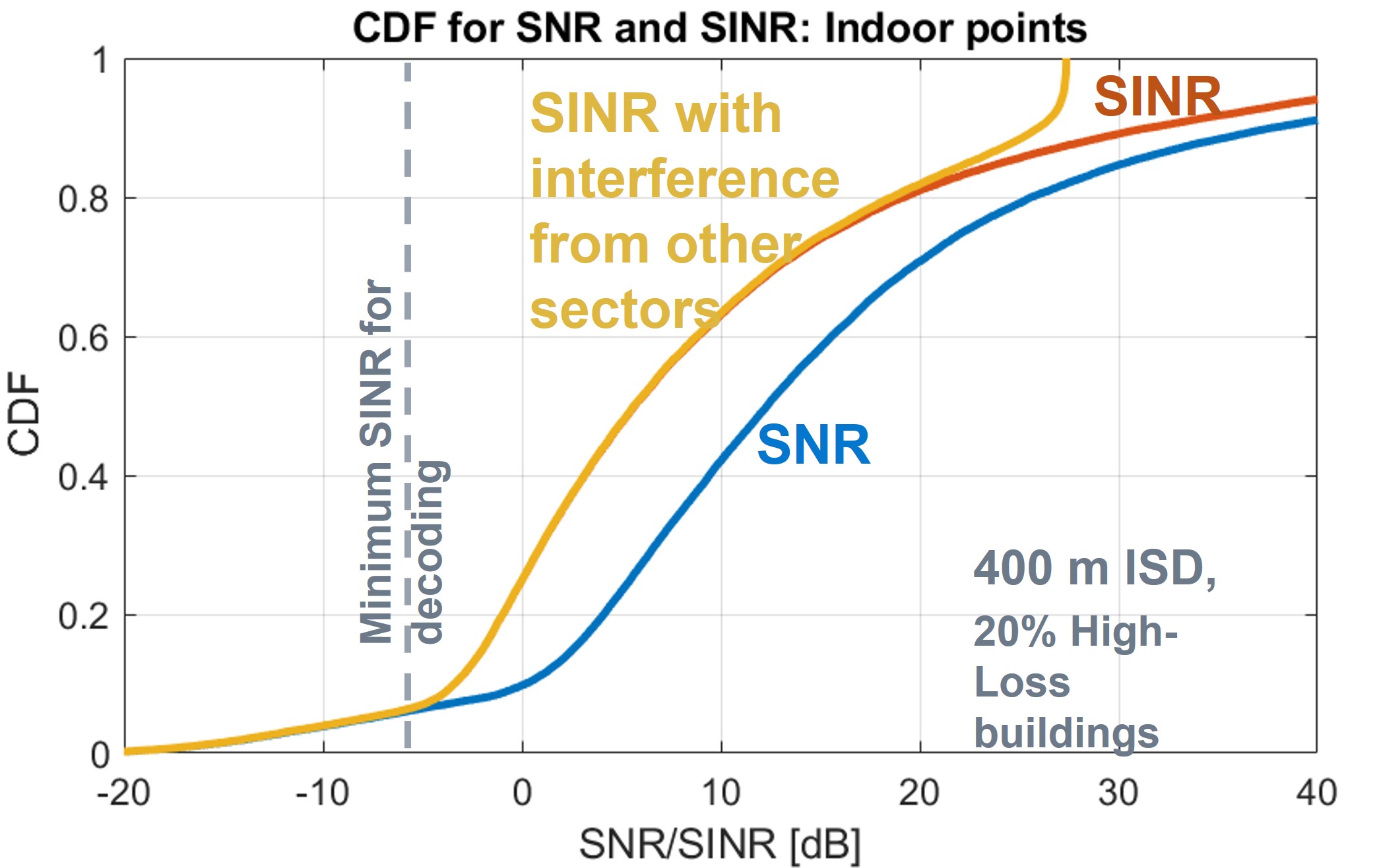}
		\caption{CDF of indoor DL SNR, SINR, and SINR considering interference from sectors at 28 GHz and transmit power of 30 dBm/pol. (Table \ref{tab:Sim_param}).}
		\label{fig:S28R}
		\vspace{-1.5 em}
	\end{figure}

	The downlink SNR coverage at 28 GHz for a time-instant is shown in Fig.~\ref{fig:heat}. The CDFs of downlink SNR and SINR  corresponding to the red-diamond boundary in Fig.~\ref{fig:Layout} are plotted in Fig.~\ref{fig:S28R}; Considering the interference from other sectors in a four-sector implementation, the SINR is limited to $\sim$27 dB SINR.  For this deployment, any user below the minimum SINR threshold of -6 dB (including a 3 dB implementation penalty) is deemed in outage \cite{ghosh2011cup}. 

	The CDF of downlink SINR for 28 GHz with 400 m ISD is shown in Fig.~\ref{fig:S28R} with one-fifth of all buildings having high-loss exteriors ($\sim$ 38 dB BPL). From the heat map in Fig.~\ref{fig:heat}, it is evident that the high loss buildings comprise the majority of the 15\% outage locations at 400 m ISD (12 BSs/sq.km). 
	A stark contrast is evident when all the buildings in the grid are considered to impose high-loss at 400 m ISD, which would have kept 61\% of the indoor users in outage. Similarly, when all buildings are low-loss ($\sim$ 18 dB BPL) at 400 m ISD only 8\% indoor users experience outage.
	
	An example of an outage location in a low-loss building is a location 10 m inside from an E-W aligned street. This might correspond to coordinate A(490 m, 490 m) on the grid, close to the edge of the 200 m diamond bound, as indicated in Fig.~\ref{fig:Layout}. The closest outdoor street location is on the E-W aligned street at (490 m, 500 m) with a 190 m Manhattan distance \cite{DuD2020tap} from the center BS at (400 m, 400 m). Using \eqref{eq:crnrPG} the path loss experienced, as the signal arrives to (490 m, 500 m), is $\approx$ 135 dB. Using the parameters specified in Table~\ref{tab:Sim_param}, an SNR of $\approx$ -13 dB is achieved indoor at point A, much lower than the -3 dB required for detection. Thus, such locations inside the low-loss buildings that are closer to E-W aligned streets with no LOS path to a BS are likely in outage.
	
\begin{figure}[t]
		\centering
		\includegraphics[width=0.46\textwidth]{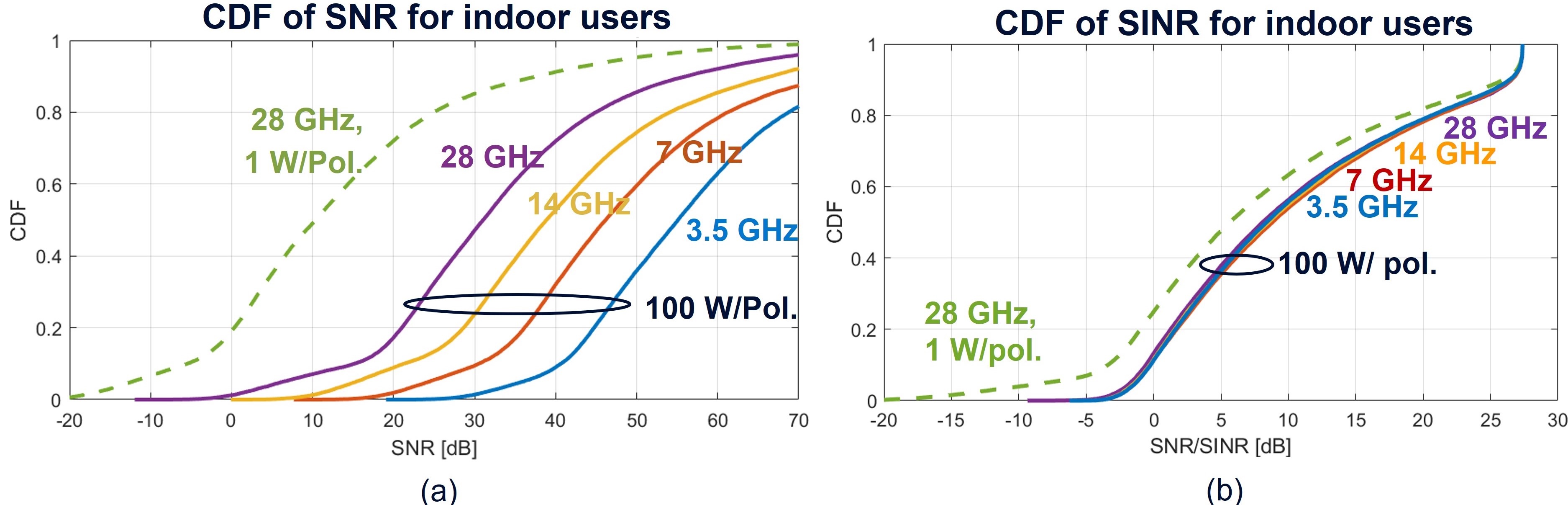}
		\caption{(a) CDF of the indoor DL SNRs at 28, 14, 7, and 3.5 GHz for 100 W/pol. transmit power and at 28 GHz for 1 W/pol. transmit power (dashed-green line).(b) CDFs of SINR with sector interference considered limits the maximum SINR achieved for the deployment at 17 dB. }
		\label{fig:S4f}
		\vspace{-0.5 em}
	\end{figure}

	Next, a direct comparison is made with lower carrier frequencies at 14, 7, and 3.5 GHz. For all frequencies the same parameter set in Table~\ref{tab:Sim_param} is used, only changing $f_c$ and increasing $P_{tx}$ to 100 W/pol. as higher transmit powers are commonly used at the lower frequency BSs \cite{Morley2019isbe}. With all parameters the same, SNR increases as center frequency decreases, as shown in Fig.~\ref{fig:S4f}(a). However, when considering interference from the other BSs and sectors, the SINR achieved is interference-limited and yields similar data rates as the SINR CDFs saturate, as captured in Fig.~\ref{fig:S4f}(b).
	Fig.~\ref{fig:S4f}(b) also shows that using 1 W/pol. power for 28 GHz, consistent with current base station products, results in a 2-3 dB degradation in SINR, an indication of being noise rather than interference limited.
	
\begin{figure}[t]
		\centering
		\includegraphics[width=0.47\textwidth]{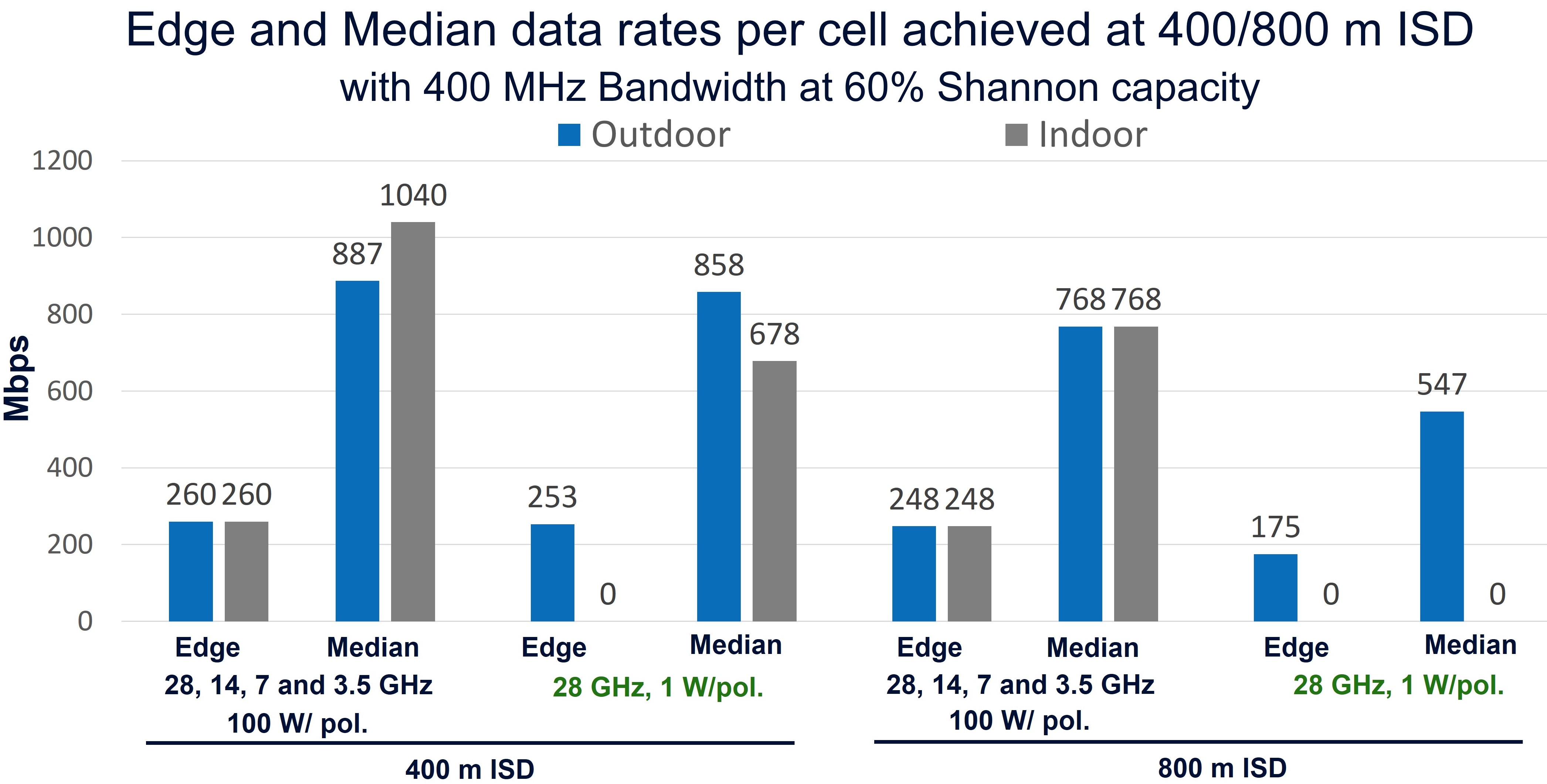}
		\caption{Similar DL data rates per cell achieved at 3.5, 7, 14, and 28 GHz with 100 W/pol.  over 400 MHz bandwidth and 80\% of all buildings low-loss at 400 m and 800 m ISD. 28 GHz with 1 W/pol. transmit power, however,  would lead to zero indoor edge rate (15\% in outage).}
		\label{fig:OIrates}
		\vspace{-1.5 em}
	\end{figure}

	The achievable downlink data rates in Mbps, calculated using Shannon capacity with 40\% overhead and 3 dB implementation penalty 
	\[Rate = (1-0.4)\times\text{BW}\times\text{Pol.}\times\log_2(1 + 10^{0.1\times(\text{SINR}-3)}), \]
	are shown in Fig. \ref{fig:OIrates}. It charts the edge (10th percentile) and median (50th percentile) downlink data rates achieved for the different $f_c$, $P_{tx}$, and ISD implemented. At the realistic power level of 1W/ pol and 28 GHz, 15\% of the indoor users are found to be in outage (400 m ISD, 20\% high-loss buildings). 
	
	The data rates for each location are calculated based on the SINR achieved. Sufficient signal power on each path results in an interference limited deployment for frequencies in the 3.5-28 GHz range, as shown in Fig.~\ref{fig:S4f}(b). The SNR achieved outdoors is higher than the indoor SNR, as expected, for the different frequencies and transmit powers. \textcolor{black}{However, the interference from other BSs at 100 W/Pol.  and 400 m ISD reduces outdoor SINR below the indoor, resulting in a higher data-rate indoors. Indoor and Outdoor points P(450, 440) and Q(450, 450) in Fig.~\ref{fig:Layout} provide an example. Without shadow fading, the signal strength at Q ($S_{Q}$) from BS1, considering around-the-corner and over-clutter-tops NLOS propagation is -11 dBm. Similarly, signal strength at P ($S_{P}$) from BS1 is -17 dBm, considering paths in Fig. \ref{fig:Paths} (i.e. $S_{P}$ indoor $< S_{Q}$ outdoor). Now, BSs 2-5 in Fig.\ref{fig:Layout} collectively provide stronger interference outdoor (-27 dBm) than indoor (-38 dBm). Thus, SINR at P indoor (-17-(-38)$ = 21$ dB) is higher than Q outdoor (-11-(-27)$ = 16$ dB). }
	
	\textcolor{black}{Increasing the ISD to 800 m weakens the interference at each location and results comparable outdoor and indoor rates as SINR CDFs converge \cite{Andrews2016cm}}.
	%
	%
	Furthermore, at 3.5 GHz, reducing transmission bandwidth from 400 MHz to 100 MHz, the edge data rate achieved at 800 m ISD was 73 Mbps and median rate obtained was 280 Mbps, comparable to \cite{Morley2019isbe}.   
	
	\section{Conclusion}
	In an urban street grid with rooftop mounted base stations at 28 GHz with 1 W transmit power per polarization, 20\% buildings imposing high loss, and 400 m ISD yielded:
	\begin{itemize}
		\item Outdoor data rate over 250 Mbps for 90\% users.
		\item 15\% of the indoor users in outage with majority of such locations within high-loss buildings.
		\item Low-loss buildings along streets without LOS path to BS and at the edge of a BS’s coverage contribute to outage.
		\item With all buildings high loss: 61\% users in outage.
	\end{itemize}
	
	At lower carriers (3.5, 7, and 14 GHz) and with 100 W/pol. power, 400 MHz bandwidth, ISD of 800 m yielded:
	\begin{itemize}
		\item A data rate of over 248 Mbps for 90\% users.
		\item A more realistic bandwidth of 100 MHz at 3.5 GHz, resulted edge rates of 73 Mbps for 3GPP Release 16. 
	\end{itemize}
	
	Thus, building-types, transmit power, and separation of base stations play a significant role in providing reliable data rate performance in dense urban scenarios.
	
	\bibliographystyle{IEEEtran}
	\bibliography{references}
	
\end{document}